\documentclass[a4paper]{jpconf}
\usepackage{graphicx}

\begin{document}
\title{AGB yields and Galactic Chemical Evolution: last updated}

\author{S. Bisterzo$^{1,2}$, C. Travaglio$^{2}$, M. Wiescher$^{3}$, R. Gallino$^{1}$, F.
K{\"a}ppeler$^{4}$, O. Straniero$^{5}$, S. Cristallo$^{5}$, G. Imbriani$^{6}$,
J. G{\"o}rres$^{3}$ and R. J. deBoer$^{3}$}
\address{$^{1}$ Dipartimento di Fisica, Universit{\`a} di Torino, Italy}
\address{$^{2}$ INAF - Astrophysical Observatory Turin, Turin, Italy}
\address{$^{3}$ Department of Physics, University of Notre Dame, Notre Dame, IN}
\address{$^{4}$ Karlsruhe Institute of Technology, Karlsruhe, Germany}
\address{$^{5}$ Osservatorio Astronomico di Collurania, Teramo, Italy}
\address{$^{6}$ Dipartimento di Scienze Fisiche, Universit{\`a} di Napoli Federico II, Italy}

\ead{bisterzo@to.infn.it,sarabisterzo@gmail.com}

%\FullConference{Nuclear Physics in Astrophysics,\\
%		May 19-24, 2013\\
%		Lisbon, Portugal}

\begin{abstract}
We study the $s$-process abundances at the epoch of the Solar-system formation
as the outcome of nucleosynthesis occurring in AGB stars of various masses and
metallicities. The calculations have been performed with the Galactic chemical 
evolution (GCE) model presented by Travaglio et al. (1999, 2004).
With respect to previous works, we used updated solar meteoritic abundances, a
neutron capture cross section network that includes the most recent measurements,
and we implemented the $s$-process yields with an extended range of AGB initial masses.
The new set of AGB yields includes a new evaluation
of the $^{22}$Ne($\alpha$, n)$^{25}$Mg rate, which takes into account the most recent 
experimental information. 
\end{abstract}

\section{Introduction}

The understanding of the $s$-process contribution to the isotopic abundances of heavy 
elements in the Solar System is fundamental to disentangle between several nucleosynthesis
processes that have competed during the evolution of the Milky Way. The $s$-process 
abundances observed in the Solar System are the result of a complex Galactic chemical 
evolution process, which mainly accounts for the pollution of several AGB generations with 
different initial masses and metallicities. 

It was shown that AGB stars with low initial masses ($M$ = 1.5 and 3 $M_\odot$), half 
solar metallicity, and a specific $^{13}$C-pocket choice (called ’case ST’) reproduce 
the $main$ component of the $s$-process (Arlandini et al. 1999), which synthesized about 
half of the elements from Sr to Pb. This approximation still provides strong information 
about the $s$-process contribution to isotopes in the region between $^{134}$Ba and $^{204}$Pb 
(see discussion in Bisterzo et al. 2011). 
\\
Two additional $s$-process components
should be considered to reproduce the solar abundances of light neutron capture isotopes 
up to Sr (the $weak-s$ component) and the stable isotopes at the termination point of the 
$s$-path, $^{208}$Pb and $^{209}$Bi (the $strong-s$ component).
The $weak-s$ process occurs in massive stars during core He and shell C burning 
and partly produces neutron capture isotopes lighter than A $\sim$ 90 ($^{86,87}$Sr
$\sim$10\%; lower contribution to Y and Zr isotopes; see e.g., Pignatari et al. 2010).
AGB stars with low metallicity and low initial mass synthesize about half of solar
$^{208}$Pb, the so called $strong-s$ component (see Gallino et al. 1998; 
Travaglio et al. 2001).
The $^{13}$C in the pocket is primary, which means it is directly synthesized in the star 
independently of the initial composition. Therefore, by decreasing the metallicity, the 
number of free neutrons per iron seed becomes so large to overcome the first and the 
second $s$-peaks (Sr-Y-Zr and Ba-La-Ce, respectively), to feed directly $^{208}$Pb 
(and $^{209}$Bi).

Actually, the understanding of the origin of light $s$-process isotopes Sr, Y and Zr is
more enigmatic. GCE models by Travaglio et al. (2004) found that AGB yields underestimate
the solar abundance of Sr, Y, Zr, and the solar composition of isotopes from $^{86}$Sr 
to $^{130}$Xe by about 20--30\% (including the $s$-only $^{96}$Mo, 
$^{100}$Ru, $^{104}$Pd, $^{110}$Cd, $^{116}$Sn, $^{122,123,124}$Te and $^{130}$Xe). 
The $weak-s$ process can not compensate the missing solar abundance, because it mainly produces 
isotopes up to Sr, with a negligible contribution up to $^{130}$Xe.
Spectroscopic observations of peculiar metal-poor stars showing large enhancement 
in $r$-process elements\footnote{E.g., CS 22892--052 and CS 31081--001 (see review by Sneden et al. 
2008), where the $rapid$ neutron capture process is ascribed to explosive nucleosynthesis phases
of massive stars.} suggest that $\sim$10\% of solar Sr-Y-Zr is 
due to the $r$-process.
However, both $s$- and $r$-processes are not sufficient to explain the solar observations of 
light neutron capture elements.
Travaglio et al. (2004) hypothesized the existence of an additional process of 
unknown origin, called by the authors LEPP (light element
primary process since it was supposed to be of primary origin).
Several scenarios have been recently explored, both involving the secondary $s$-process 
in massive stars (e.g., $cs$-component by Pignatari et al. 2013, which may explain
the missing $s$-process component) or primary $r$-process during the advanced 
phases of explosive nucleosynthesis (see review Thielemann et al. 2011, which may
account of complementary $r$-contributions). 
Therefore, even if promising theoretical improvements related to the explosive phases 
of massive stars and core collapse Supernovae, as well as recent spectroscopic
investigations (Roederer et al. 2012; Hansen et al. 2012) have been made,
a fully understanding about the origin of the neutron capture elements from Sr up to 
Xe is still lacking.

We aim to investigate the effects of a new set of AGB yields (with updated nuclear 
cross section network and solar abundances) on the solar $s$-process composition (Section~2).

\section{Results}

We focus the analysis on the variations of the Solar-system $s$-process GCE predictions
by adopting updated AGB yields, and by testing the effects of different prescriptions 
on nuclear cross sections. 
The Galactic evolution is computed as function of time up 
to the present epoch ($t_{Today}$ = 13.73 $\pm$ 0.12 by WMAP),
following the three zones of the Galaxy, halo, thick and thin disk. 
We adopted the yields by Rauscher et al. (2002) and Travaglio et al. (2004b) for 
SNe II and Ia, respectively.

Major revisions involve the solar system 
meteoritic abundances by Lodders, Palme and Gail (2009), a neutron capture cross 
section network with the most recent published measurements, as well as a larger 
set of AGB yields that extend toward lower initial mass (down to $M$ = 1.3 $M_\odot$).
We started from the AGB models presented by Bisterzo et al. (2010), which were based 
on the FRANEC code by Straniero et al. (1995, 2000, 2003).
\\
A series of thousands of new AGB models have been run, for a total
of 28 metallicities from [Fe/H] = +0.2 down to $-$3.6 (most of them
focused on the metallicity range between solar and [Fe/H] = $-$1.6, where the 
isotopes of the three $s$-peaks are largely produced, see Travaglio et al. 1999, 2001,
2004; Serminato et al. 2009).
Yields of a set of five AGB models with low initial masses ($M$ = 1.3, 1.4, 1.5, 2, 
and 3 $M_\odot$) and two AGB models with intermediate initial masses ($M$ = 5 and 
7 $M_\odot$) have been interpolated/extrapolated over the whole metallicity and mass 
range (hereafter LMS refers to the mass range between 1.3 $\le$ $M/M_\odot$ $<$ 4,
and IMS to 4 $\le$ $M/M_\odot$ $<$ 8). 
This assures a sufficiently high accuracy in the AGB mass and metallicity ranges. 

Particularly challenging for the $s$-process is the understanding of the 
formation of the $^{13}$C-pocket, specifically the mass fraction of $^{13}$C 
and $^{14}$N in the pocket and the mass involved. 
Both uncertainties largely affect the $s$-yields (see e.g., Straniero et al. 2006; 
Herwig 2005). 
Current full evolutionary AGB models still adopt a free parametrization to 
reproduce the $^{13}$C-pocket, by means of overshooting (e.g., Herwig 2000, Herwig et
al. 2011; Karakas 2010) or other better physically justified prescriptions (e.g., 
new FRUITY models by Cristallo et al. 2009, 2011, and references therein; or 
mixing produced by magnetic fields, Busso et al. 2012). 
As suggested by the $s$-process spread observed at a given metallicity in different stellar populations 
(e.g., post-AGB, Ba, CH and CEMP-s stars; Sneden et al. 2008; K{\"a}ppeler et al 2011), 
a range of the $s$-process efficiency strengths is needed. 
The theoretical reason of this spread observed among $s$-elements 
is still under investigation (AGB initial mass, 
magnetic fields, gravitational waves, or rotation, see Piersanti, Cristallo and
Straniero 2013).
\\
As discussed by Bisterzo et al. (2010), we artificially introduce 
the $^{13}$C-pocket in our post-process AGB calculations, and we treated it as a free 
parameter kept constant pulse by pulse. 
Starting from the $^{13}$C-pocket ST case, similar to that adopted by Gallino et al. (1998), 
we multiply or divide the $^{13}$C (and $^{14}$N) abundances in the pocket by different 
factors.
We considered an accurate weighted
average of the $^{13}$C-pocket efficiencies
in order to reproduce $\sim$100\% of solar $^{150}$Sm and the other $s$-only 
isotopes heavier than $A$ $\sim$ 90.
Note that $^{150}$Sm has well defined solar abundance, and it is the 
$s$-only isotope less affected by branchings
and nuclear cross section uncertainties as well.
A second reaction, the $^{22}$Ne($\alpha$, n)$^{25}$Mg, starts to partially burn 
at $T_8$ = 3 during thermal pulses 
and produces an efficient neutron burst 
mainly affecting isotopes close to the branching points.

\begin{figure}[h]
\includegraphics[angle=-90,width=35pc]{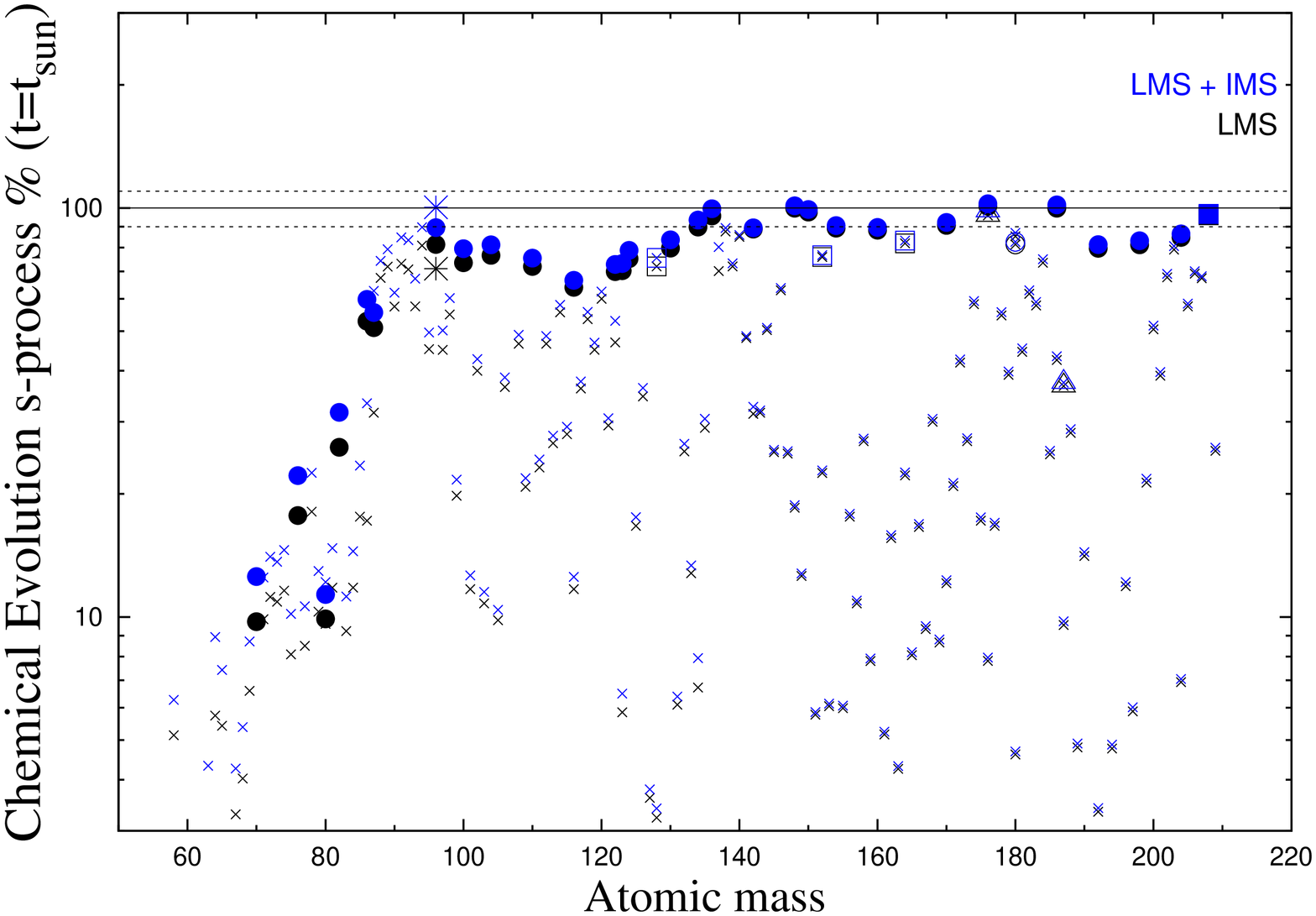}\hspace{2pc} 
\includegraphics[angle=-90,width=35pc]{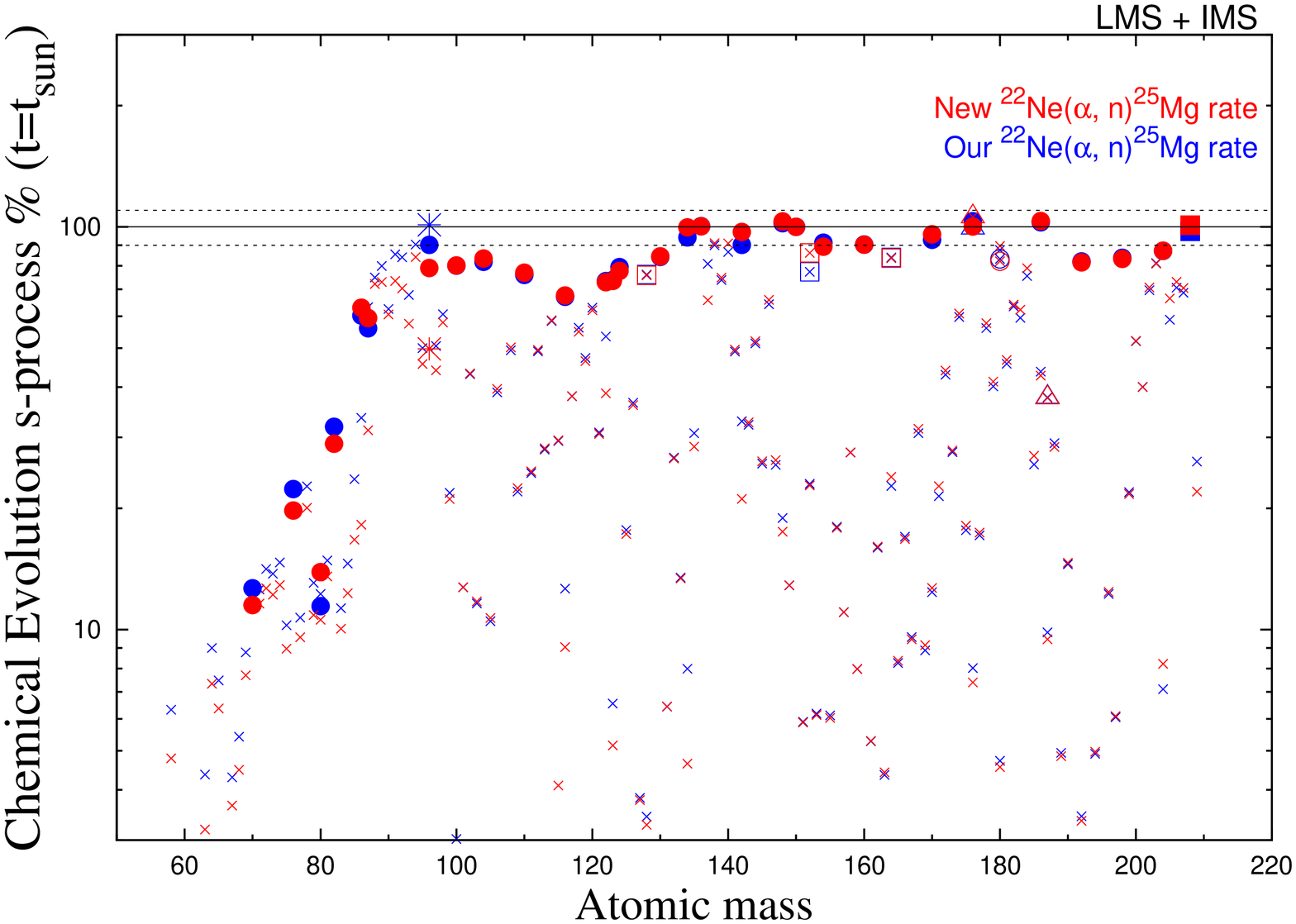}\hspace{2pc} 
\caption{\label{Fig1} Reproduction of the Solar $s$-process abundances (in \%)  
obtained at the epoch of the Solar System formation with GCE model. 
We used a neutron capture cross section network that includes the most recent
measurements (see Bisterzo et al. in preparation) and we implemented the s-process 
yields with an extended range of AGB initial masses.
Updated solar meteoritic abundances by Lodders et al. (2009) are adopted.
The $s$-only isotopes are indicated by solid circles. 
$^{96}$Zr is represented by a big asterisk. Different symbols have been 
used for isotopes that receive additional contributions (see text).
$Top$ $panel$: we distinguish between the contribution of LMS alone (black symbols)
and the total $s$-contribution of all AGB masses (LMS + IMS; blue symbols).
$Bottom$ $panel$: we display the total $s$-contribution shown in top panel (blue) in comparison to 
the results obtained with a new evaluation of the $^{22}$Ne($\alpha$, 
n)$^{25}$Mg reaction (see text).}
\end{figure}

Results are shown in Figure 1. The $s$-only isotopes are indicated by solid circles. 
Different symbols have been 
used for isotopes that receive additional contributions: $^{128}$Xe, $^{152}$Gd, 
and $^{164}$Er (open squares), which have a non-negligible p contribution;
$^{176}$Lu (open triangle), a long-lived isotope (3.8 x 10$^{10}$ yr), which decays 
into $^{176}$Hf; $^{187}$Os (open triangle), which is affected by the long-lived 
decay of $^{187}$Re (4.1 x 10$^{10}$ yr); $^{180}$Ta (open circle), which also receives 
contributions from the p process and from $\nu$-nucleus interactions in massive stars; 
$^{208}$Pb (filled big square), half of which is produced by the strong-$s$ component.
We compare the $s$-process predictions, calculated with the GCE model at the epoch of 
the Solar System formation ($t_{\odot}$ = 9.17 Gyr from the beginning of the birth of
the Universe), with the meteoritic abundances by Lodders et al. (2009).
As shown in the $top$ $panel$, LMS (black symbols) 
reproduce almost all the Solar System $s$-only isotopes between $A$ = 140 and 210.
An additional small contribution ($<$10\%) comes from IMS for isotopes with $A$ $<$ 140. 
The total $s$-percentages (LMS + IMS) is represented by blue symbols.
IMS AGBs play a minor role in the Galactic enrichment, because their He-intershell is smaller 
than in LMS by one order of magnitude, with an uncertain formation of the $^{13}$C-pocket
and a less efficient third dredge-up. 
The $^{22}$Ne($\alpha$, n)$^{25}$Mg neutron source is efficiently activated in IMS due 
to the high temperature reached at the bottom of the thermal pulses ($T_8$ = 3.5).
The peak neutron 
density reached in IMS easily allows a strong overproduction of $^{86}$Kr, $^{87}$Rb, and 
$^{96}$Zr, three neutron rich isotopes affected by the branchings at $^{85}$Kr and $^{95}$Zr,
which are very sensitive to the neutron density. 
\\
Particularly large is the abundance of $^{96}$Zr (big asterisk).
Note that $^{96}$Zr is strongly sensitive to the number of thermal pulses
experienced by the AGB models, as well as to the $^{22}$Ne($\alpha$, n)$^{25}$Mg rate adopted. 
In general, an over-prediction of $^{96}$Zr may suggest a low number of thermal pulses
for IMS. As discussed by Bisterzo et al. (2010), 
we assume a strong mass loss for IMS, which allows a total of 24 thermal pulses. 
Under this hypothesis, IMS produce about 30\% of solar $^{96}$Zr, while an additional 70\% 
comes from LMS (Figure 1, $top$ $panel$).
$^{96}$Zr is the most neutron rich stable Zr isotope, and
a contribution of 100\% from AGB  
is surely overestimated.
Moreover, it disagrees with recent $p$-process predictions by Travaglio et al. (2011), which
estimate an additional non-negligible contribution to $^{96}$Zr by SNIa (up to 30\%).

As a further improvement, we computed a new set of AGB yields that includes a new evaluation
of the $^{22}$Ne($\alpha$, n)$^{25}$Mg rate, which takes into account the most recent 
experimental information. The recommended value we are suggesting is very close to Jaeger
et al. (2001) and agrees with the recent determination of Longland et al. (2012).
At AGB temperature ($T_8$ $\sim$ 2.5 to 3.5) the new $^{22}$Ne($\alpha$, n)$^{25}$Mg rate is about 
a factor of 2 lower than our rate used so far (corresponding to the lower limit suggested by
K{\"a}ppeler et al. 1994)\footnote{Note that we modified accordingly the rate of the 
$^{22}$Ne($\alpha$, $\gamma$)$^{26}$Mg. Specifically, at the temperature of interest of AGBs, 
between $T_8$ = 2.5 to 3.5, this rate is almost unchanged.}.
The related new Solar-system abundances are displayed in Figure 1, $bottom$ $panel$, red symbols.
As expected, major differences are shown close to the branchings points.
In particular, $\sim$50\% of solar $^{96}$Zr is produced by AGBs, in better agreement
with expectations.
\\
Updated GCE calculations plausibly reproduce within the uncertainties all $s$-only isotopes 
with $A$ $>$ 130, and confirm the missing 20\% Solar $s$-contribution of $s$-only isotopic 
abundances between $A$ = 96 -- 130 found by Travaglio et al. (2004).
Variations with respect to the results presented by K{\"a}ppeler et al. (2011; their Fig.~15, 
bottom panel) are mainly due to new solar abundances, recent neutron capture cross section
measurements, and the new evaluation of the $^{22}$Ne($\alpha$, n)$^{25}$Mg rate.
\\
Note that $^{192}$Pt and $^{198}$Hg are affected by large uncertainties:
concerning $^{192}$Pt, the neutron capture cross section of $^{191}$Os and $^{192}$Ir 
evaluated theoretically 
at 22\%, the extrapolation of the $^{192}$Ir measurement in stellar conditions (see 
discussion by Rauscher et al. 2012), as well as the old measurement of the $^{192}$Pt(n, 
$\gamma$) reaction, with 20\% of uncertainty at 30 keV (Bao et al. 2000); 
moreover, Hg is too volatile for a reliable experimental determination of
the solar abundance (Lodders et al. 2009 estimated an uncertainty of 20\%).
$^{204}$Pb (and all Pb isotopes) have well determined neutron capture cross sections
(see KADoNiS), but it is strongly affected by the branching at $^{204}$Tl, with variations 
of $\sim$10\%.

Updated Solar $s$-process abundances of some selected elements are compared with
previous GCE computations (Travaglio et al. 2001, 2004) in Table~\ref{tab1}.
Marginal differences ($<$5\%) are seen in general.
La and Ce are among the few exceptions: the larger $s$-contribution obtained by
this work (+12\% and +6\%) is the consequence of the new $^{139}$La(n, $\gamma$)$^{140}$La
rate measured by Winckler et al. (2006).
A more detailed comparison and a discussion concerning the most relevant updated
information will be provided in a future work.

\begin{table}[h]
\caption{\label{tab1} The $s$-process contribution from LMS and IMS AGBs at the epoch
of the Solar System formation (in percentages).
Results by Travaglio et al. (2001, 2004) in column 2 are compared with 
updated results (this work, column 3), computed with all recently updated information,
including the new $^{22}$Ne($\alpha$, n)$^{25}$Mg rate. The $r$-process contribution
(evaluated with the residual method $N_r$ = $N_s$ -- $N_\odot$)
is given in column 4 (in percentages).} 
\begin{tabular*}{5cm}{*{13}{c}}
%\br 
  &  &  &   &  &  & & &  &      &     &   &  \cr
  &  &  &   &  &  & & &  &   El   &   s(2001-2004) & s(This work) & r(This work) \cr
%\mr       
  &  &  &   &  &  & & &  &   Sr   & 71  & 67&  ---   \cr
  &  &  &   &  &  & & &  &   Y    & 69  & 70&  ---   \cr
  &  &  &   &  &  & & &  &   Zr   & 65  & 64&  ---   \cr
  &  &  &   &  &  & & &  &   Ba   & 80  & 83&  17  \cr
  &  &  &   &  &  & & &  &   La   & 61  & 73&  27  \cr
  &  &  &   &  &  & & &  &   Ce   & 75  & 81&  19  \cr
  &  &  &   &  &  & & &  &   Pr   & 47  & 49&  51  \cr
  &  &  &   &  &  & & &  &   Nd   & 54  & 56&  44  \cr
  &  &  &   &  &  & & &  &   Sm   & 30  & 31&  69  \cr
  &  &  &   &  &  & & &  &   Eu   & 6   &  6&  94  \cr
  &  &  &   &  &  & & &  &   Pb   & 91  & 88&  12  \cr
%\br
\end{tabular*}
\end{table}

\ack 

S.B. wishing to acknowledge JINA for financial support under ND Fund
$\#$201387 and 305387.

\end{document}